\documentclass[preprints,communication,accept,moreauthors,pdftex]{Definitions/mdpi}
\firstpage{1} 
\makeatletter 
\pubvolume{9}
\issuenum{1}
\articlenumber{0}
\pubyear{2023}
\copyrightyear{2023}
\hreflink{https://doi.org/} 
\pdfoutput=1

\makeatletter
\let\c@lofdepth\relax
\let\c@lotdepth\relax
\makeatother

\usepackage[final]{changes}  

\usepackage{subfigure}
\makeatletter
\renewcommand{\@thesubfigure}{\normalsize(\textbf{\alph{subfigure}})}
\makeatother
\Title{Do Multi-Structural One-Off FRBs Trace Similar Cosmology History with Repeaters?}

\TitleCitation{Do Multi-Structural One-Off FRBs Trace Similar Cosmology History with Repeaters?}

\Author{Yu-Hao Zhu
~$^{1,2}$\orcidA{}, Chen-Hui Niu~$^{3,}$*\orcidB{}, Xiang-Han Cui~$^{1,2}$\orcidC{}, Di Li~$^{1,2,4,}$*\orcidD{}, Yi Feng~$^{5}$\orcidE{}, Chao-Wei Tsai~$^{1}$\orcidF{}, Pei~Wang~$^{1}$\orcidG{}, Yong-Kun Zhang~$^{1,2}$\orcidH{}, Fanyi Meng~$^{2}$\orcidI{} and Zheng Zheng $^{1}$}

\AuthorCitation{Zhu, Y.-H.; Niu, C.-H.; Cui, X.-H.; Li, D.; Feng, Y.; Tsai, C.-W.; Wang, P.; Zhang, Y.-K.; Meng, F.Y.; Zheng, Z.}

\address{%
$^{1}$ \quad National Astronomical Observatories, Chinese Academy of Sciences, Beijing 100101, China; \linebreak zhuyh@bao.ac.cn (Y.-H.Z.);  cuixianghan@nao.cas.cn (X.-H.C.); cwtsai@nao.cas.cn (C.-W.T.); wangpei@nao.cas.cn (P.W.); ykzhang@nao.cas.cn (Y.-K.Z.); zz@bao.ac.cn (Z.Z.)\\
$^{2}$ \quad University of Chinese Academy of Sciences, Beijing 100049, China; mengfanyi@ucas.ac.cn\\
$^{3}$ \quad Institute of Astrophysics, Central China Normal University, Wuhan 430079, China\\
$^{4}$ \quad NAOC-UKZN Computational Astrophysics Centre, University of KwaZulu-Natal, Durban 4000, South Africa\\
$^{5}$ \quad Research Center for Intelligent Computing Platforms, Zhejiang Laboratory, Hangzhou 311100, China; yifeng@zhejianglab.com\\
}

\corres{Correspondence: peterniu@nao.cas.cn (C.-H.N.); dili@nao.cas.cn (D.L.)}

\abstract{Fast Radio Bursts (FRBs) are millisecond-duration transient events that are typically observed at radio wavelengths and cosmological distances but their origin remains unclear. Furthermore, most FRB origin models are related to the processes at stellar scales, involving neutron stars, blackholes, supernovae, etc. In this paper, our purpose is to determine whether multi-structural one-off FRBs and repeaters share similarities. To achieve this, we focus on analyzing the relationship between the FRB event rate and the star formation rate, complemented by statistical testing methods.
Based on the CHIME/FRB Catalog 1, we calculate the energy functions for four subsamples, including apparent non-repeating FRBs (one-offs), repeaters, multi-structural one-offs, and the joint repeaters and multi-structural events, respectively. We then derive the FRB event rates at different redshifts for all four subsamples, all of which were found to share a similar cosmological evolution trend. However, we find that the multi-structural one-offs and repeaters are distinguishable from the KS and MWW tests.}


\begin{document}

\section{Introduction}
\label{sec:intro}

The field of FRB research has undergone a rapid development in terms of observational analysis, leading to significant progress in understanding this enigmatic astronomical event~\cite{Chatterjee 2017,chime second repeater,Caleb 2021 progress,Petroff 2019 review,Zhang 2020}. Since the first fast radio burst (FRB) was published in 2007~\cite{Lorimer 2007}, more than 600 FRBs have been detected~\cite{Petroff FRBCAT 2016,chime 2021}. The Canadian Hydrogen Intensity Mapping Experiment (CHIME) largely enhanced the detected FRB samples, which made it possible to study FRB populations more precisely~\cite{chime 2021, Chen 2022}. {Meanwhile, numerous studies have investigated the potential connections between FRBs and GRBs, as well as the cosmological aspects related to FRBs (e.g.,~\cite{Deng 2014, Gao 2014, Zhou 2014, Madhavacheril 2019}).} Despite the increasing number of detected FRBs, their origins and radiation mechanisms remain uncertain. To explain the possible origins of FRBs, numerous models have been proposed~\cite{Cordes 2019,Petroff 2021,Lyubarsky 2021}. Platts et al.~\cite{Platts 2019} provided a detailed summary of FRB radiation mechanisms and potential progenitors. Zhang et al.~\cite{Zhang 2020} also summarized a catalog of possible models to explain the origins of FRBs. Among these models, Li et~al.~\cite{Li 2021} confirmed that some FRBs likely originate from magnetars but it is still unclear if magnetars are the origins of all FRBs~\cite{Lin 2020, chime 2020 magnetar, Bochenek 2020}.

Hashimoto et al.~\cite{Hashimoto 2020 no evolution} analyzed the event rates of non-repeating FRBs and repeaters and found that they trace different cosmological histories. Locatelli et al.~\cite{Locatelli 2019} and James et~al.~\cite{James 2021} discovered that FRBs exhibit redshift evolution behavior. Arcus et al.~\cite{Arcus 2021 dm distribution sfr} used a method similar to that of James et al.~\cite{James 2021} to find that both the star formation rate density and stellar mass density evolutionary histories can describe the FRB evolutionary history. Zhang et al.~\cite{Zhang 2020 redshift distribution}, using Parkes and ASKAP FRB samples, also found that they could not distinguish between the two evolutionary histories.

Generally, FRBs are usually divided into two categories: repeating and non-repeating FRBs based on whether repeating bursts have been observed. Most of the current FRB samples are non-repeating FRBs. Although repeating FRBs and non-repeating FRBs have different widths, bandwidths~\cite{Morphology 2021}, rotation measures~\cite{Feng 2022}, and variation time-scales of the circular polarization~\cite{Feng 2022b}, whether all FRB sources repeat is still in debate~\cite{Caleb 2019 n/r, Cui 2020}. To determine if an FRB is a repeater, one must spend plenty of time consistently monitoring an FRB source~\cite{Caleb 2021 progress, Connor 2018 r/nr}. Monitoring all FRBs requires a significant time investment, yet expanding the sample size of repeating bursts is crucial for understanding their origins. As the number of observations grows, some non-repeating FRBs display multi-component frequency-time patterns reminiscent of those observed in repeating FRBs~\cite{Morphology 2021, Hessels 2019}. Consequently, we attempt to identify that some non-repeating FRBs with these multi-structures could potentially be candidates for repeating FRBs {\cite{Hessels 2019}}. Energy functions are essential to calculate the FRB number density and allow one to investigate the redshift evolution of volumetric occurrence rates of FRBs~\cite{Hashimoto 2020 no evolution, Hashimoto 2022 energy functions, James 2021}. Our purpose is to enhance our comprehension of FRB origins by examining the relations of multi-structural one-offs, repeaters, and non-repeating FRBs. To achieve this, we investigate their event rate evolution and statistical~properties.

The structure of this paper is as follows: Section \ref{sec:data selection} elaborates on the data selection criteria, FRB samples utilized in this study, and the categorization of FRBs. In Section \ref{sec:method}, we provide an in-depth explanation of the research methodologies implemented in this paper. Section \ref{sec:results} presents the findings derived from energy functions, event rates, and statistical tests. By comparing the results from both event rate analyses and statistical tests, we draw our conclusions in Section \ref{sec:conclusion}.

\section{Data Selection}
\label{sec:data selection}

\subsection{Selection Function}
\label{sec:selection function}

To avoid a sample bias in the FRB population analysis, it is essential to consider the sample bias caused by the instrument selection effect to obtain a relatively accurate sample distribution. CHIME collaboration~\cite{chime 2021} developed a method for calculating the selection function of CHIME through simulated FRBs. Based on the real FRBs detected in CHIME/FRB Catalog~1, Hashimoto et al.~\cite{Hashimoto 2022 energy functions} provided the selection function for FRBs with a signal-to-noise(SNR) greater than 10. In this study, we use the selection functions (5)--(8) from Hashimoto et~al.~\cite{Hashimoto 2022 energy functions}. By implementing the selection functions, we can rectify the FRB number density, thereby circumventing the influence of instrument effects on the computation of event~rates.

\subsection{Samples}
\label{sec:samples}

This paper utilizes data from the CHIME/FRB Catalog 1. As the first FRB survey catalog released by CHIME/FRB, it encompasses 536 bursts detected between 25 July 2018 and 1 July 2019. The survey spans 214.8 days and includes 474 non-repeating FRBs and 18 repeaters~\cite{chime 2021}. The availability of the CHIME/FRB Catalog 1 facilitates more precise statistical analysis in FRB population studies. While composing this paper, CHIME identified 25 new repeating FRBs~\cite{chime 2023 25 new}. To maximize the sample size without introducing new biases, we cross-referenced these 25 newly discovered repeating bursts with the CHIME/FRB Catalog 1. We found that some one-off FRBs in the CHIME/FRB Catalog 1 exhibited repeating characteristics. Thus, we updated the information for these newly added repeating bursts in the CHIME/FRB Catalog 1, which increased the number of repeaters and slightly reduced the number of non-repeating FRBs accordingly. In order to ensure the data used reflects the intrinsic energy distribution of FRBs as accurately as possible, we first filtered the CHIME sample using the following selection criteria: \newpage

\begin{enumerate}
    \item bonsai\_snr > 10;
    \item $\mathrm{DM_{obs}} > 1.5 \times max(\mathrm{DM_{NE2001}}, \mathrm{DM_{YMW16}})$;
    \item Not detected in far side-lobes;
    \item $\mathrm{log\,} \tau_\mathrm{scat} < 0.8$ (ms);
    \item excluded\_flag = 0;
    \item The first detected burst if the FRB source is a repeater;
    \item $\mathrm{log}\, F_{\nu} > 0.5 $ (Jy ms).
\end{enumerate}

After applying the above selection function, we retained 202 FRBs, including 17 repeating FRBs and 185 one-offs. Within these 185 one-offs, we found that some FRBs have multiple pulse components (multi-structures). After manual examination, we identified 18 multi-structural one-offs in the current sample. To conduct a statistical analysis and event rate analysis of this subset, we analyzed them as a distinct category. The detailed sample classification can be found in Section \ref{sec:data group}.

\subsection{Classification}
\label{sec:data group}

In order to study the relationship between the event rates and redshifts of different types of FRBs, we divided the FRB samples into four groups:

\begin{enumerate}
    \item One-offs without multi-structures (167 cases);
    \item Repeaters (20 cases);
    \item Multi-structural one-offs (18 cases);
    \item Repeaters + multi-structural one-offs (38 cases).
\end{enumerate}

In the process of identifying multi-structural one-offs, we employed the criteria of being able to distinguish two or more pulse profiles in the time domain. Ultimately, we selected 18 FRBs from one-offs in CHIME/FRB Catalog 1 which satisfied the selection functions as the sample of multi-structural one-offs.

The purpose of dividing the data into the above four categories was to explore whether different classes of FRBs share similar evolution with redshifts and to investigate the possibility that multi-structural one-offs could be candidates for repeaters. In the group of one-offs, to avoid contamination of the non-repeating FRBs sample by multi-structural one-offs, we removed FRBs with multi-structures from the one-offs.

\section{Method}
\label{sec:method}

\subsection{Redshift Calculation}
\label{sec:redshift-DM}

In the sample we used, only a portion of the FRBs had redshifts derived from optical localization. Moreover, Michilli et al.~\cite{Michilli 2022} reported 13 FRBs with baseband localization and we cross-referenced these 13 FRBs with redshift measurements with those in the CHIME/FRB Catalog 1, updating the redshifts for our sample accordingly. For FRBs with optical redshifts, we used the redshifts obtained from optical instead of redshifts calculated from the DM-z relation for subsequent calculations. For other FRBs without optical redshifts, we employed dispersion measures (DMs) to determine their corresponding redshifts.

The observed DM of FRBs is composed of the following components:
\begin{equation}
\label{equ:DM_obs}
\mathrm{DM_{obs}} = \mathrm{DM_{MW}} + \mathrm{DM_{halo}} + \mathrm{DM_{IGM}}(z) + \mathrm{DM_{host}}(z),
\end{equation}
where $\mathrm{DM_{MW}}$ represents the Galactic dispersion contribution, for which we employed the $\mathrm{DM_{MW,YMW16}}$ model for estimation. To account for the redshift variation caused by $\mathrm{DM_{host}}$, we adopted a log-normal distribution for $\mathrm{DM_{host}}$. In our work, we adopted the mean value $\mu=\mathrm{\ln(68.2)}$ and the variance $\sigma_\mathrm{host}=0.88$ in the log-normal distribution, reported by Macquart et al.~\cite{Macquart 2020 cosomo}. Afterwards, we divided the $\mathrm{DM_{host}}$ obtained from the log-normal distribution by a factor of $(1+z)$, yielding the final contribution of $\mathrm{DM_{host}}$. Following Hashimoto et al.~\cite{Hashimoto 2022 energy functions}, we utilized $\mathrm{DM_{halo}}=65\, \mathrm{pc\, cm^{-3}}$. The calculation of $\mathrm{DM_{IGM}}$ refers to Equation (2), (4) in Maquart et al.~\cite{Macquart 2020 cosomo}, and the cosmological parameters used in this paper are from the Planck15 cosmological model. {Deriving the redshift through $\mathrm{DM}$ usually relies on $\mathrm{DM_{IGM}}$ and its fluctuation. The fluctuation of $\mathrm{DM_{IGM}}$ causes a large error in the estimated redshift. In this work, we used Equations (B1) and (B2) derived by Hashimoto et~al.~\cite{Hashimoto 2020 no evolution} as the $\mathrm{DM_{IGM}}$ fluctuation function.} Based on Equation (\ref{equ:DM_obs}), we can directly calculate the redshift using the Markov Chain Monte Carlo (MCMC) method through DM.

\subsection{Calculation of Energy Functions}
\label{sec:ef}

After obtaining the redshifts of FRBs, we can estimate the luminosity distances of the FRBs based on their redshifts and convert the observed energy to the energy in the rest frame, enabling us to examine the energy distributions of FRBs.

To study the event rate as a function of redshift, we need to calculate the energy function in different redshift intervals. Therefore, for the four groups of FRBs mentioned in Section \ref{sec:data group}, we first divided the redshift intervals as follows: the redshift intervals of one-offs are $z=$ 0.05, 0.30, 0.68, 1.38, 3.60; due to the relatively small redshifts of repeaters and multi-structural one-offs, we decided to use the redshift intervals of $z=$ 0.05, 0.31, 0.72, and 1.50. Subsequently, following Equations (10)--(23) from Hashimoto et al.~\cite{Hashimoto 2022 energy functions}, we use the Vmax method to calculate the number density of the FRBs~\cite{Schmidt 1968}, and based on the calculated results of the number density, we obtained data points for the energy functions in different redshift intervals, ultimately deriving the energy equations in different redshift intervals. In the calculation of the number density, we used the CHIME selection functions for correction, where our sample yielded $W_\text{scale}=0.928$.

\subsection{Statistical Tests}
\label{sec:statistical}

In order to analyze whether the three groups of FRBs in Section \ref{sec:data group} have similar distributions from another perspective, we conducted the Kolmogorov–Smirnov (K–S) test~\cite{ks} on the observed parameters of the three groups of FRBs to determine if the parameter distributions of each group exhibit significant differences. To avoid test biases caused by insufficient sample sizes from repeaters and multi-structural one-offs as much as possible, we also took the Mann–Whitney–Wilcoxon test (M–W–W) and the Wilcoxon rank sum test~\cite{mww} on the observed parameters of the FRB groups.

The K–S test, M–W–W test, and Wilcoxon rank-sum test are all non-parametric tests. Compared to the K–S test, the M–W–W test and Wilcoxon rank-sum test are more suitable for statistical tests with small samples~\cite{mww}. Each statistical test provides two parameters: statistical correlations and p-values. Generally, when the p-value is below 0.05, the two test samples are considered to come from different distributions. In this work, the test threshold (significance level) we employed is $5 \%$. Table \ref{tab:ks}, \ref{tab:mww}, \ref{tab:rs} are the results of K–S test, M-W-W test, and Wilcoxon rank-sum test, respectively.

\begin{table}[H]
\tablesize{\footnotesize}
    \caption{\emph{p}-values from K–S test.}
    \label{tab:ks}
    \newcolumntype{C}{>{\centering\arraybackslash}X}
    \begin{tabularx}{\textwidth}{CCCCCCCCC}
    \toprule
        ~ & \textbf{\boldmath{$\mathrm{DM}_\text{exc,ymw16}$}} & \textbf{Width} & \textbf{\boldmath{$\tau_\text{scat}$}} & \textbf{Fluence} & \textbf{\boldmath{$\Delta \nu_{\mathrm{FRB}}$}} & \textbf{\boldmath{$E_\text{rest,400}$}}   \\ 
        ~ & \textbf{(\boldmath{$\mathrm{pc\, cm^{-3}}$})} & \textbf{(ms)} & \textbf{(ms)} & \textbf{(Jy ms)} & \textbf{(MHz)} & \textbf{(erg)}\\
        \midrule
        Test 1 & 0.087  & 0.011  & 0.099  & 0.606  & $3.502~\times~10^{-5}$   & 0.618 \\ 
        Test 2 & 0.104  & 0.970  & 0.365  & 0.242  & 0.229   & 0.656  \\ 
        Test 3 & 0.781  & {0.034}  & 0.053  & 0.135  & 0.058   & 0.676  \\ 
    \bottomrule
    \end{tabularx}
    \noindent{\footnotesize{Test 1 is statistical test between one-offs and repeaters, test 2 is statistical test between one-offs and multi-structural one-offs, and test 3 is statistical test between repeaters and multi-structural one-offs.}}
\end{table}
\unskip

\begin{table}[H]
    \caption{\emph{p}-values from M–W–W test.}
    \label{tab:mww}
    \newcolumntype{C}{>{\centering\arraybackslash}X}
    \begin{tabularx}{\textwidth}{CCCCCCC}
    \toprule
        ~ & \textbf{\boldmath{$\mathrm{DM}_\text{exc,ymw16}$}} & \textbf{Width} & \textbf{\boldmath{$\tau_\text{scat}$}} & \textbf{Fluence} & \textbf{\boldmath{$\Delta \nu_{\mathrm{FRB}}$}} & \textbf{\boldmath{$E_\text{rest,400}$}}   \\ 
        ~ & \textbf{(\boldmath{$\mathrm{pc\, cm^{-3}}$})} & \textbf{(ms)} & \textbf{(ms)} & \textbf{(Jy ms)} & \textbf{(MHz)} & \textbf{(erg)}\\
        \midrule
        Test 1 & 0.028  & 0.002  & 0.109  & 0.901  & $1.913~\times~10^{-5}$  & 0.177   \\ 
        Test 2 & 0.078  & 0.711  & 0.518  & 0.150  & 0.130   & 0.638   \\ 
        Test 3 & 0.304  & {0.006}  & {0.016}  & 0.070  & {0.021}   & 0.265   \\ 
    \bottomrule
    \end{tabularx}
    \noindent{\footnotesize{Test 1 is statistical test between one-offs and repeaters, test 2 is statistical test between one-offs and multi-structural one-offs, and {test 3} is statistical test {between repeaters and multi-structural one-offs}.}}
\end{table}

\begin{table}[H]
    \caption{\emph{p}-values from Wilcoxon rank-sum test.}
    \label{tab:rs}
    \newcolumntype{C}{>{\centering\arraybackslash}X}
    \begin{tabularx}{\textwidth}{CCCCCCCCC}
    \toprule
        ~ & \textbf{\boldmath{$\mathrm{DM}_\text{exc,ymw16}$}} & \textbf{Width} & \textbf{\boldmath{$\tau_\text{scat}$}} & \textbf{Fluence} & \textbf{\boldmath{$\Delta \nu_{\mathrm{FRB}}$}} & \textbf{\boldmath{$E_\text{rest,400}$}}   \\ 
        ~ & \textbf{(\boldmath{$\mathrm{pc\, cm^{-3}}$})} & \textbf{(ms)} & \textbf{(ms)} & \textbf{(Jy ms)} & \textbf{(MHz)} & \textbf{(erg)}\\
        \midrule
        Test 1 & 0.028  & 0.002  & 0.108  & 0.899  & $4.529~\times~10^{-5}$   & 0.177   \\ 
        Test 2 & 0.078  & 0.709  & 0.517  & 0.149  & 0.152 & 0.637   \\ 
        Test 3 & 0.599  & {0.011}  & {0.032}  & 0.136  & {0.041}    & 0.520   \\ 
    \bottomrule
    \end{tabularx}
    \noindent{\footnotesize{Test 1 is statistical test between one-offs and repeaters, test 2 is statistical test between one-offs and multi-structural one-offs, and test 3 is statistical test between repeaters and multi-structural one-offs.}}
\end{table}

\section{Results and Discussion}
\label{sec:results}

\subsection{Energy Functions}
\label{sec:ef_result}

We found that, in the lower redshift interval, the energy function of one-offs exhibits a steeper slope at higher energies and a relatively flatter slope at lower energies, resembling the Schechter energy function shape. Consequently, we took the Schechter energy function (Luo et al.~\cite{Luo 2018 lf}) to fit the energy function for the four subsamples of FRBs. {The Schechter energy function is characterized by:}

\begin{figure}[H]
    
    \includegraphics[width=13.5cm]{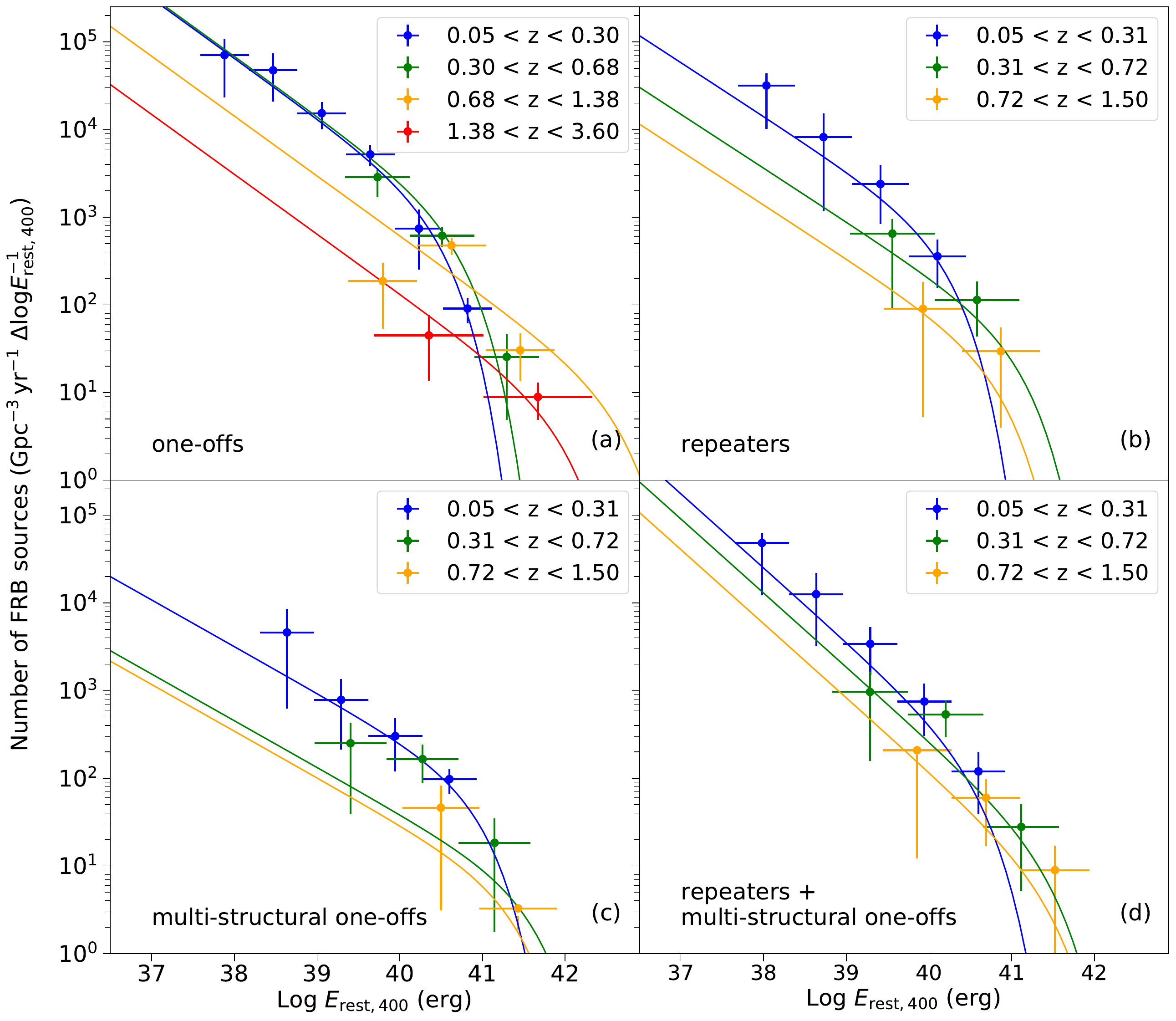}
    \caption{Energy 
 functions of four FRB groups; (\textbf{a})--(\textbf{d}) demonstrate 
 the energy functions of one-offs, repeaters, multi-structural one-offs, and the merged group, repeaters, and multi-structural one-offs as a whole, respectively. The vertical errors are given by MCMC. The lines are the best-fitted~result.}
    \label{fig:efs}
\end{figure}

{
\begin{equation}
    \label{equ:schechter function}
    \phi(\log E) \mathrm{d} \log E=\phi^*\left(\frac{E}{E^*}\right)^{\alpha+1} \exp \left(-\frac{E}{E^*}\right) \mathrm{d} \log E,
\end{equation}
where $\phi^*$ is the normalization factor, $\alpha$ is the slope in the linear scale, and $E^*$ is the cut-off energy.
} When fitting the energy functions, since there are relatively more data in the first redshift interval, we used the first redshift interval to fit the slope of the energy functions. The slope of the remaining redshift intervals was determined using the fitting results of the first redshift interval and the remaining parameters were fitted. Finally, the best-fitting values are shown in Table 
 \ref{tab:fitting_result}. The best-fitted Schechter energy functions are demonstrated in Figure \ref{fig:efs}.

\begin{table}[H]
    \caption{Best-fitted 
 parameters of FRB energy functions.}
    \label{tab:fitting_result}
    \newcolumntype{C}{>{\centering\arraybackslash}X}
    \begin{tabularx}{\textwidth}{CCCC}
        \toprule
       \textbf{ Redshift Bin }& \boldmath{$\phi^*$} & \boldmath{$E_\text{\textbf{rest,400}}^*$} & \boldmath{$\alpha$} \\
        \midrule
        \multicolumn{4}{c}{One-offs} \\
        $0.05 < z \leq 0.30$ & $3.14_{-0.37}^{+0.30}$ & $40.46_{-0.14}^{+0.20}$ & $-1.68_{-0.14}^{+0.13}$\\
        $0.30 < z \leq 0.68$ & $2.99_{-0.20}^{+0.28}$ & $40.70_{-0.29}^{+0.19}$ & $-1.68 ^{a}$\\
        $0.68 < z \leq 1.38$ & $0.97_{-0.48}^{+0.60}$ & $42.66_{-1.00}^{+0.91}$ & $-1.68 ^{a}$\\
        $1.38 < z \leq 3.60$ & $0.77_{-0.56}^{+1.84}$ & $41.99_{-2.80}^{+1.04}$ & $-1.68 ^{a}$\\
        \multicolumn{4}{c}{Repeaters} \\
        $0.05 < z \leq 0.31$ & $2.80_{-0.23}^{+0.28}$ & $40.19_{-0.14}^{+0.26}$ & $-1.61_{-0.18}^{+0.24}$\\
        $0.31 < z \leq 0.72$ & $1.66_{-0.47}^{+0.59}$ & $41.09_{-0.69}^{+0.95}$ & $-1.61 ^{a}$\\
        $0.72 < z \leq 1.50$ & $1.39_{-0.27}^{+0.46}$ & $40.85_{-0.58}^{+0.68}$ & $-1.61 ^{a}$\\
        \multicolumn{4}{c}{Multi-structural one-offs} \\
        $0.05 < z \leq 0.31$ & $1.94_{-0.50}^{+0.44}$ & $40.94_{-0.41}^{+0.39}$ & $-1.53_{-0.29}^{+0.32}$\\
        $0.31 < z \leq 0.72$ & $0.74_{-0.48}^{+0.64}$ & $41.59_{-0.88}^{+0.79}$ & $-1.53 ^{a}$\\
        $0.72 < z \leq 1.50$ & $0.73_{-0.54}^{+0.67}$ & $41.39_{-0.62}^{+0.54}$ & $-1.53 ^{a}$\\
        \multicolumn{4}{c}{Repeaters + multi-structural one-offs} \\
        $0.05 < z \leq 0.31$ & $2.22_{-0.94}^{+1.01}$ & $40.57_{-0.80}^{+0.66}$ & $-1.84_{-0.26}^{+0.33}$\\
        $0.31 < z \leq 0.72$ & $1.18_{-0.76}^{+0.75}$ & $41.47_{-0.75}^{+0.93}$ & $-1.84^{a}$\\
        $0.72 < z \leq 1.50$ & $0.81_{-0.56}^{+0.79}$ & $41.50_{-0.96}^{+0.84}$ & $-1.84^{a}$\\
        \bottomrule
    \end{tabularx}
     \noindent{\footnotesize{Note: $a$ represents a fixed value.  $E_\text{rest,400}^*$ is the cut-off energy in the Schechter function.}}
\end{table}
We discovered that, within the error range, the slope of the energy equation for one-offs in our result is basically consistent with the results of James et al.~\cite{James 2021}{.} Due to the limitation of th$ $e sample size of repeaters and multi-structural one-offs, we can only provide a rough constraint on the slope of energy functions, respectively. We estimate that a larger sample size would allow us to obtain more accurate constraints.

\subsection{History of Redshift Evolution}
\label{sec:event rate}

The calculation of FRB event rates is an effective approach to understand the nature of FRBs~\cite{Hashimoto 2022 energy functions}. To investigate whether different types of FRBs have similar evolutions and to explain whether multi-structural one-offs may have similarities to repeaters, we calculate the relationship between event rates and redshift for each group of FRBs.

\begin{figure}[H]
    
    \includegraphics[width=13.5cm]{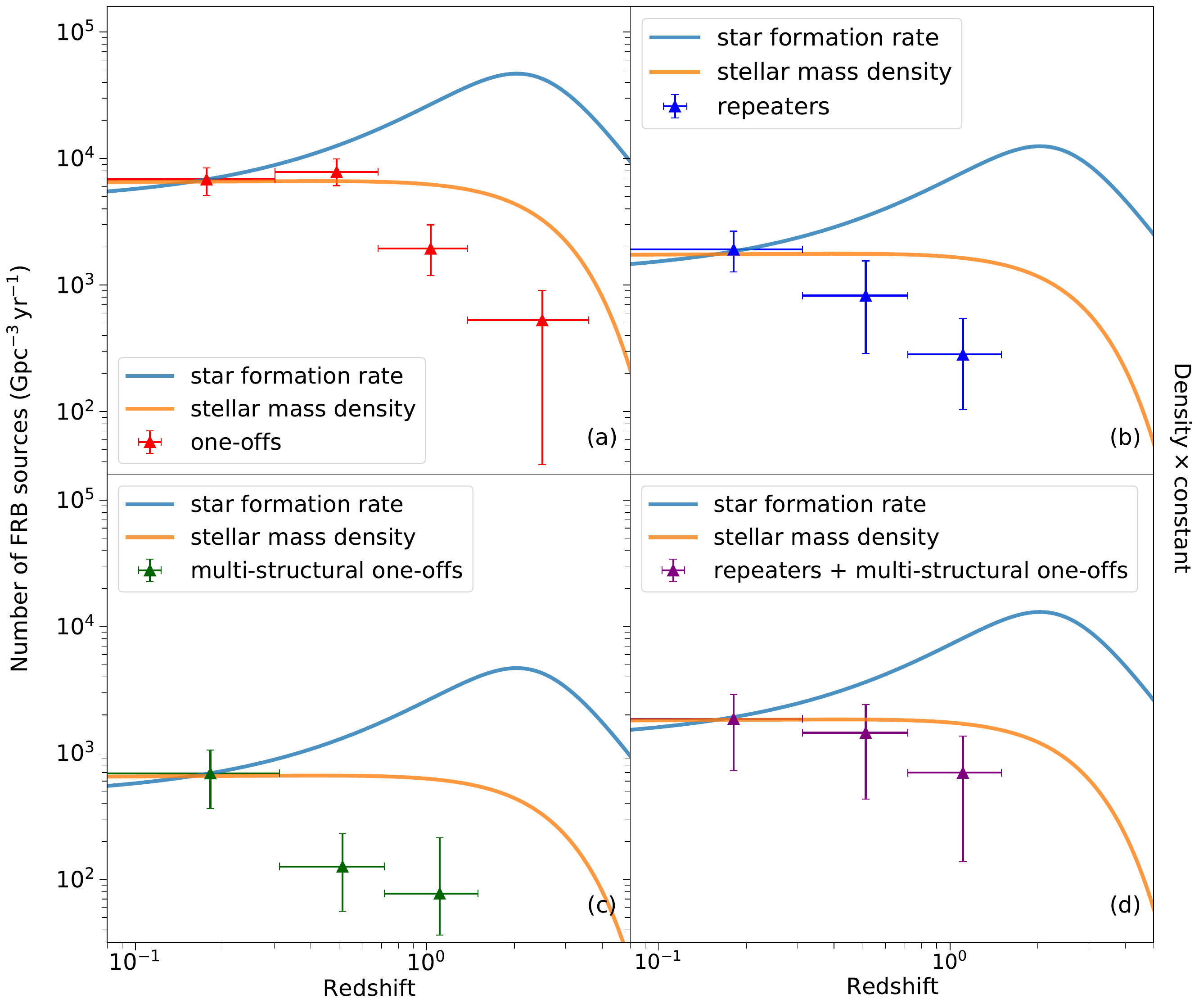}
    \caption{Event rate relation with redshift; (\textbf{a})--(\textbf{d}) demonstrate the event rate of one-offs, repeaters, multi-structural one-off, and the merged group, repeaters, and multi-structural one-offs as a whole, respectively. The blue and yellow lines stand for cosmic star-formation rate and cosmic stellar-mass density evolution history, respectively. The vertical errors in the volumetric event rates are given by MCMC from energy function fitting. To compare the event rate trend of different groups of FRBs with redshift, we adjusted the star formation rate and stellar mass density to be same as the number of sources at the first redshift value in each group by a multiple of the constant to the density.}  

    \label{fig:event_rates}
\end{figure}

By integrating the energy functions corresponding to the four groups of FRBs in Section \ref{sec:ef_result} at different redshift intervals, we can obtain the event rates of each FRB group at different redshifts. The uncertainties in the event rates are directly calculated using the MCMC method during the energy function fitting process. The energy integration range we selected is $(10^{39},\, 10^{41.5})\, \text{erg}$, consistent with the ranges used in Shin et al.~\cite{Shin 2023} and Hashimoto et al.~\cite{Hashimoto 2020 no evolution, Hashimoto 2022 energy functions}. Figure \ref{fig:event_rates} demonstrates the relationships between event rates and redshifts for one-offs, repeaters, and multi-structural one-offs, respectively.

{To compare the volumetric event rate of FRBs with both cosmic star formation rate density and cosmic stellar mass density, we used the following two equations. First is the star formation rate density given by Madau et al.~\cite{Madau 2017} denoted as follows:}

{
\begin{equation}
    \label{equ:sfr}
    \log \left(\psi_{\mathrm{SFRD}}\right)=\log \left\{0.01 \frac{(1+z)^{2.6}}{1+[(1+z) / 3.2]^{6.2}}\right\}\left(\mathrm{M}_{\odot} \mathrm{yr}^{-1} \mathrm{Mpc}^{-3}\right),
\end{equation}
}
{and second is the cosmic stellar mass density from Hashimoto et al.~\cite{Hashimoto 2020 no evolution}. This function is the fitting result of the observed data in L\'opez Fern\'andez et al.~\cite{Lopez 2018}:}

{
\begin{equation}
    \label{equ:stellar mass density}
    \begin{aligned}
        \log \left(\rho_*\right)= & 8.156+5.906 \times 10^{-2} z-7.111 \times 10^{-2} z^2 \\
        & +4.034 \times 10^{-2} z^3-1.256 \times 10^{-2} z^4+2.209 \times 10^{-3} z^5 \\
        & -2.216 \times 10^{-4} z^6+1.179 \times 10^{-5} z^7 \\
        & -2.585 \times 10^{-7} z^8\left(\mathrm{M}_{\odot} \mathrm{Mpc}^{-3}\right).
    \end{aligned}
\end{equation}
}

The relationship between event rates and redshifts indicates that, within the $1\sigma$ uncertainty, the cosmological history of non-repeating FRBs is inconsistent with the evolutionary history of the star formation rate. This conclusion aligns with Hashimoto et al.~\cite{Hashimoto 2022 energy functions}. Based on the current sample size, we believe that the relationship between event rates and redshift for the four groups of FRBs tends to align more with the evolutionary history of stellar mass density.

\subsection{Statistical Analysis}
\label{sec:statistic_results}

The results of the K–S test from Table \ref{tab:ks} demonstrate that there are noteworthy disparities in the distributions of dispersion, pulse width, and bandwidth between repeaters and non-repeating FRBs, indicating that they may have different distributions. Under the same parameters, the distribution of repeaters and multi-structural one-offs is different.


{In the aspect of non-repeating FRBs and repeaters, }the cumulative distribution plot demonstrated in Figure \ref{fig:test}(b) illustrates the differences in pulse width distributions between repeaters and non-repeating FRBs. It is evident that, after normalizing the two distributions, there are differences in the CDF distributions. The mean value of pulse widths in the non-repeating FRBs are narrower than those of repeaters, implying that the radiation mechanisms of repeating and non-repeating FRBs may be different.

While in the aspect of multi-structural one-offs and repeaters, the results of the statistical analysis show different distributions. Tables \ref{tab:mww} and \ref{tab:rs} present the results of the M–W–W and Wilcoxon rank-sum tests, respectively. Although some parameters show differences in their test results compared to the K–S test, the overall conclusions {implied by the results} remain consistent: the origins of repeaters and {multi-structural} one-offs may be different. It is also found that the mean values of the pulse width and scattering time of multi-structural one-offs are lower than repeaters, indicating that multi-structural one-offs may have unique characteristics compared to repeaters. {Additionally,} multi-structural one-offs do not show significant differences from non-repeating FRBs. It is important to note that the statistical results may be biased due to the small sample size. Therefore, with a larger sample size for both groups, the conclusions may differ from those currently obtained.

\section{Conclusions}
\label{sec:conclusion}

{``Multi-structural'' is believed to indicate the presence of repeating FRBs~\cite{Hessels 2019}.} In this paper, we calculated the energy functions of four subsamples, apparent non-repeating FRBs, repeaters, multi-structural one-offs, and the joint sample of repeaters and multi-structural one-offs using the Vmax method in CHIME/FRB Catalog 1. After analyzing the event rate evolution history and conducting the K–S and M–W–W tests, we drew the following~conclusions:
\begin{itemize}
    \item The trend of the event rate of non-repeating FRBs is consistent with what was found by Hashimoto et al.~\cite{Hashimoto 2020 no evolution}. The redshift evolution of the event rate of repeaters, however, shows no similarity with that of the star formation rate, which was suggested by Hashimoto et al.~\cite{Hashimoto 2020 no evolution} and James et al.~\cite{James 2021}. We calculated the energy functions for 167 one-offs, 20 repeaters, 18 multi-structural one-offs, and 38 repeaters and multi-structural one-offs combined. The redshift evolution of all four subsamples was found to be distinct from that of the star formation rate.

    \item We carried out statistical testing to quantify the similarities between multi-structural one-offs and repeaters. {We found that, based on the current sample size, the difference between repeaters and multi-structural one-offs can be distinguished from statistical analysis and the differences between multi-structural one-offs and usual one-offs are not significant. Through K–S and M–W–W test, we can conclude that multi-structural one-offs may have differences to repeaters. Whether morphological multi-structural one-offs are candidates of repeaters still needs to be determined when a larger sample size is available.}
    

\end{itemize}


\vspace{6pt} 



\funding{This work was funded by the National Natural Science Foundation of China (No. 11988101, 12203069) and the National SKA Program of China/2022SKA0130100. This work was also supported by the Office the Leading Group for Cyberspace Affairs, CAS (No. CAS-WX2023PY-0102). Z.Z. is supported by NSFC grant No. 12041302, and U1931110. 
Z.Z. is also supported by the science research grant from the China Manned Space Project with grant no. CMS-CSST-2021-A08.}

\dataavailability{The dataset we used in the work is available in the CHIME/FRB Catalog 1 repository at 10 March 2023, 
 \url{https://www.chime-frb.ca/catalog}.}

\acknowledgments{The authors gratefully acknowledge valuable discussions with T. Hashimoto, and Y.C. Li. And, we greatly acknowledge two anonymous referees for their valuable comments and suggestions, which have significantly improved the quality of the paper.}

\conflictsofinterest{The authors declare no conflict of interest.} 

\appendixtitles{no} 
\appendixstart
\appendix
\section[\appendixname~\thesection]{}
Appendix A shows the observed parameter distributions of one-offs, repeating FRBs, and multi-structural one-offs, where the upper (lower) panels correspond to the cumulative (differential) distribution of the observed parameters, respectively.

\begin{figure}[H]
    \centering
    \subfigure[~DM distributions between one-offs and repeaters.]{
    \label{fig:1}
    \includegraphics[width=0.45\linewidth]{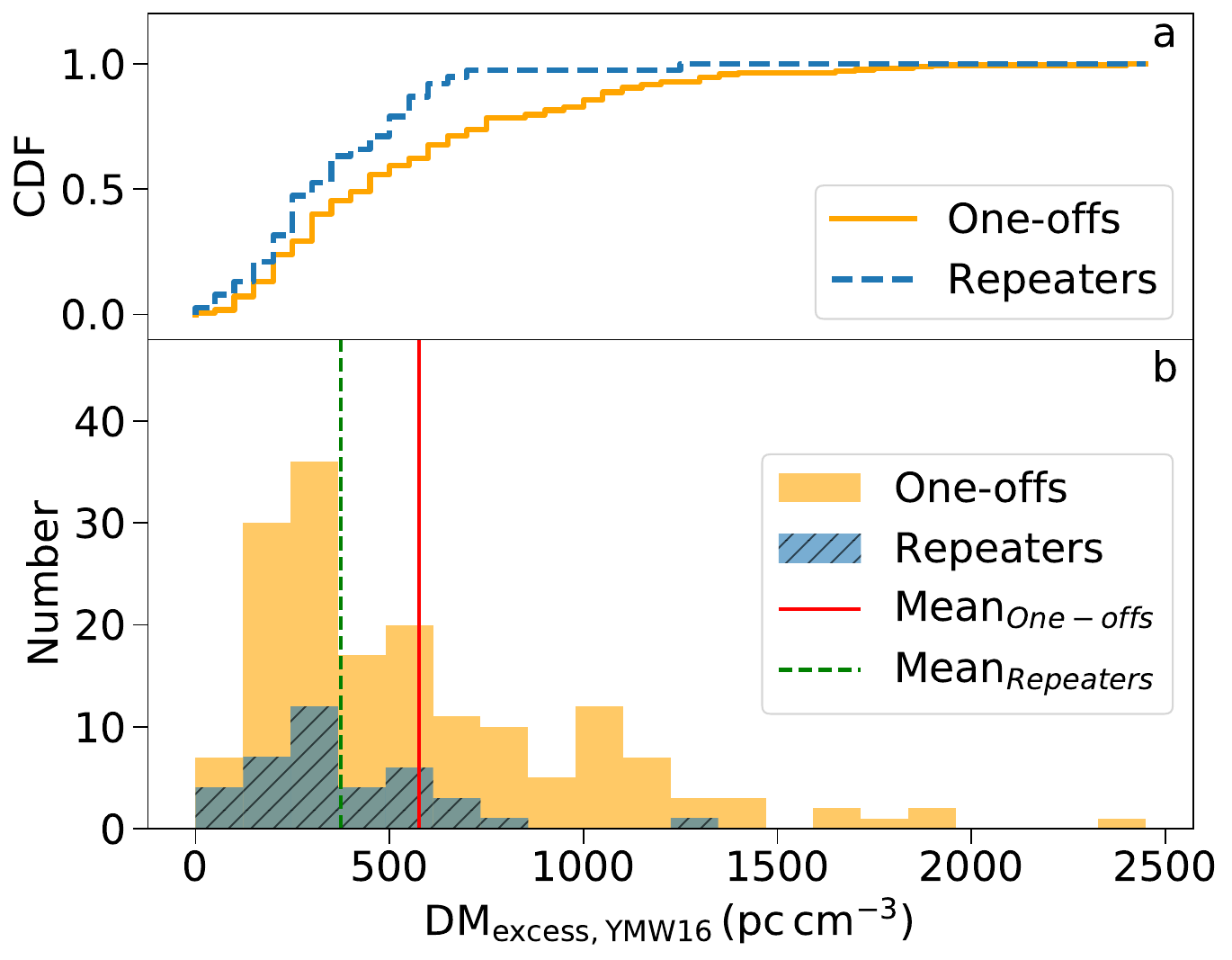}
    }
    \quad
    \subfigure[~Width distribution between one-offs and repeaters.]{
    \label{fig:2}
    \includegraphics[width=0.45\linewidth]{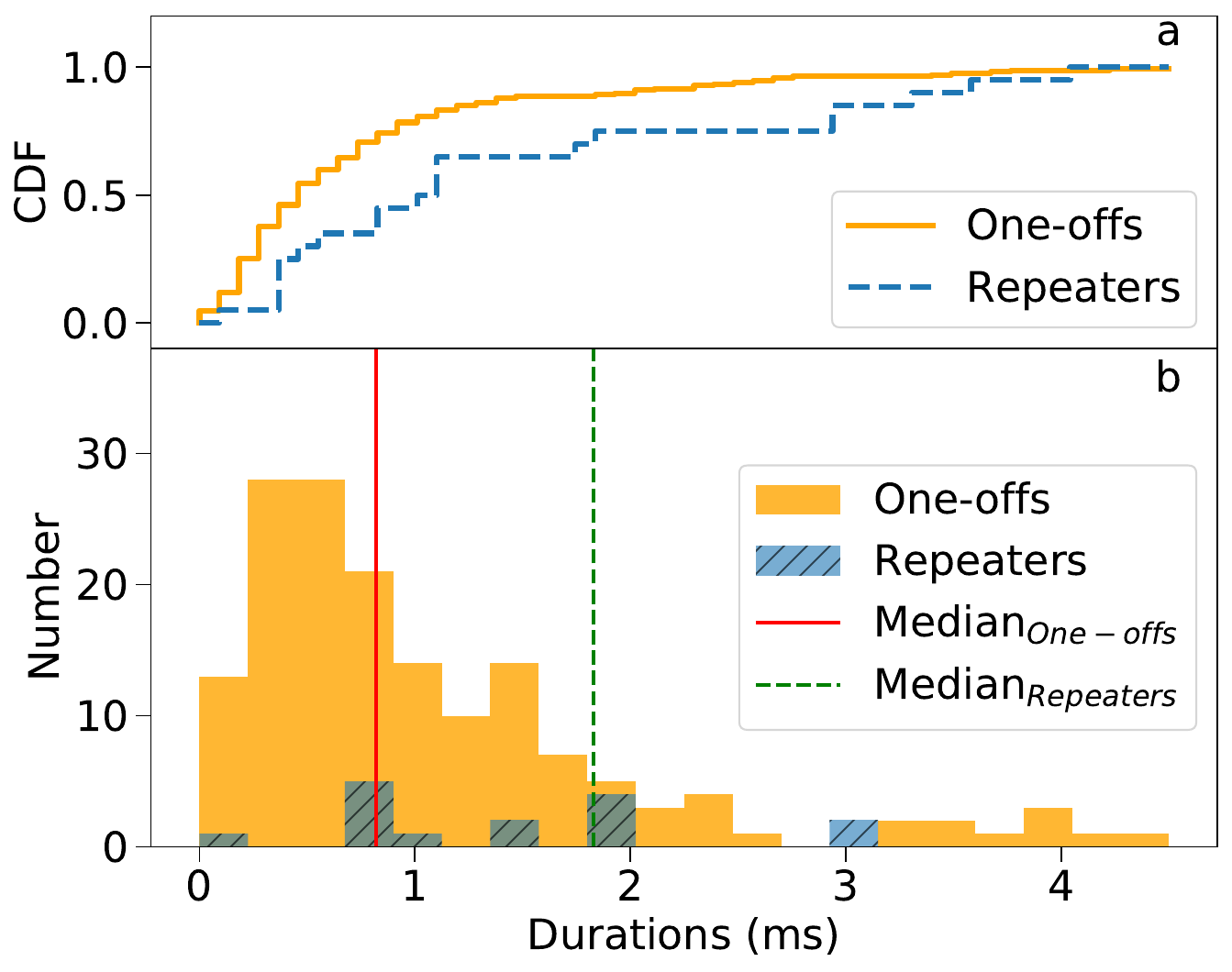}
    }
    \quad
    \subfigure[~Bandwidth distributions between one-offs and repeaters.]{
    \label{fig:3}
    \includegraphics[width=0.45\linewidth]{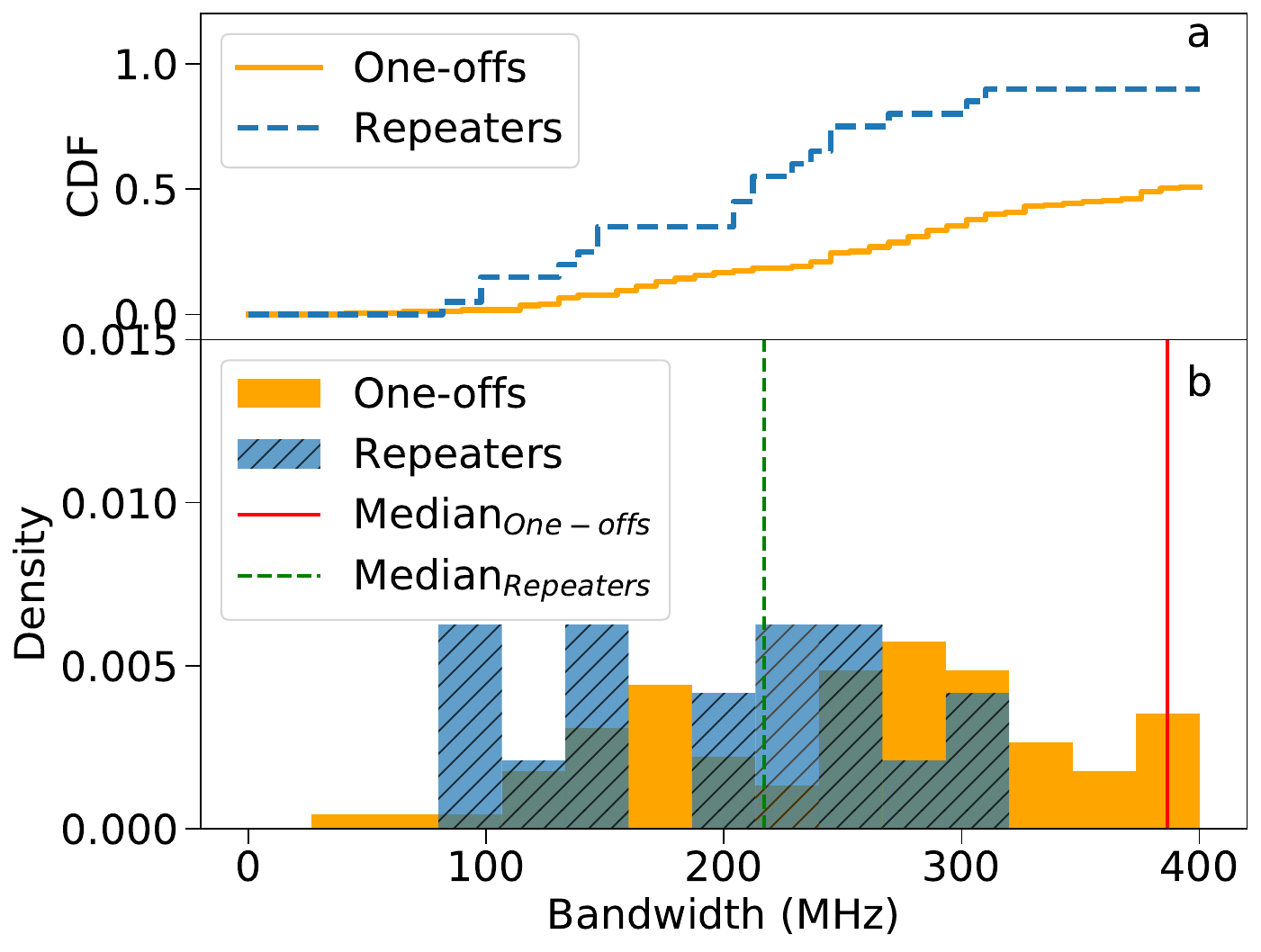}
    }
    \quad
    \subfigure[~Width distributions between repeaters and multi-structural one-offs.]{
    \label{fig:4}
    \includegraphics[width=0.45\linewidth]{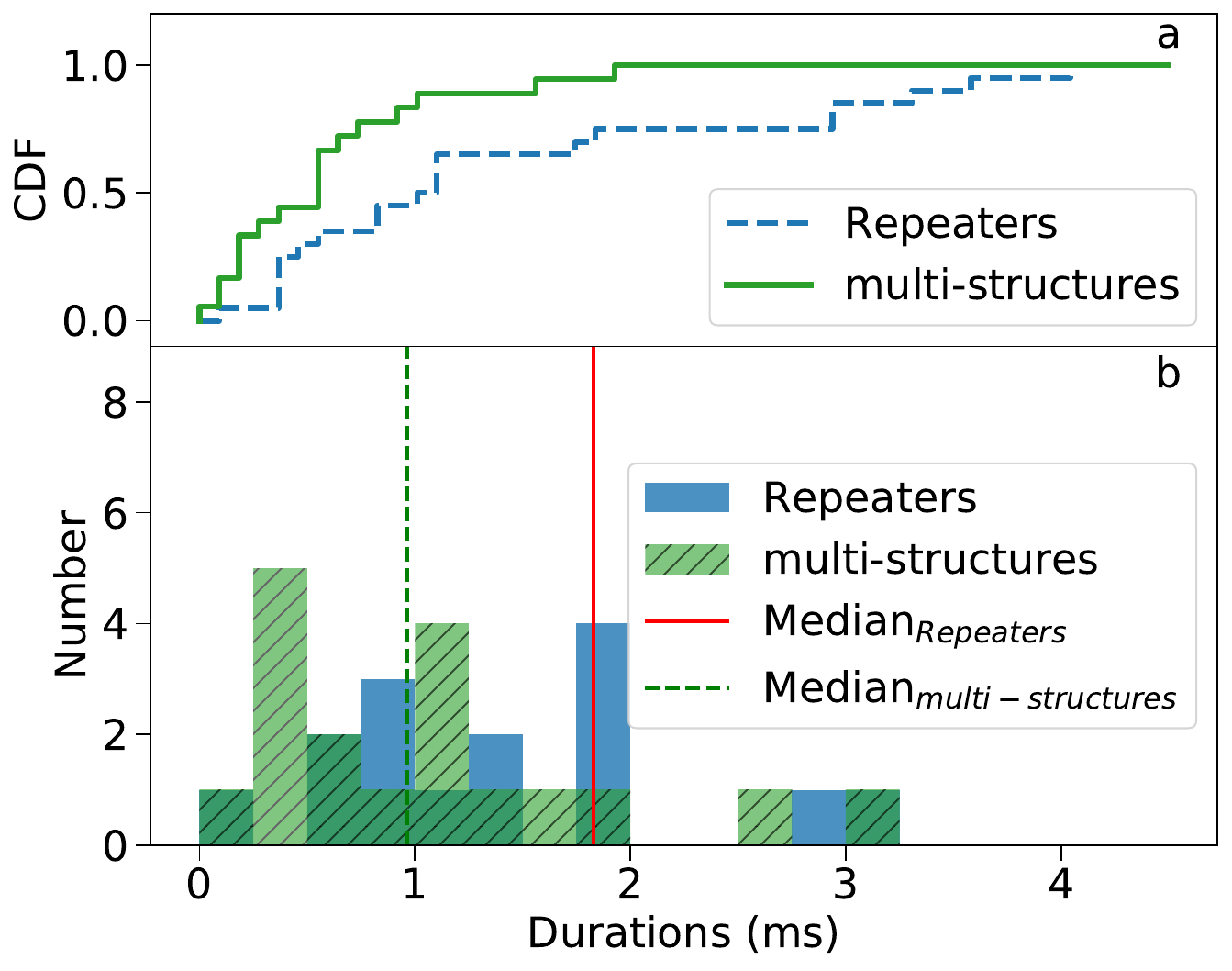}
    }
    \quad
    \subfigure[~Scattering distributions between repeaters and multi-structural one-offs.]{
    \label{fig:5}
    \includegraphics[width=0.45\linewidth]{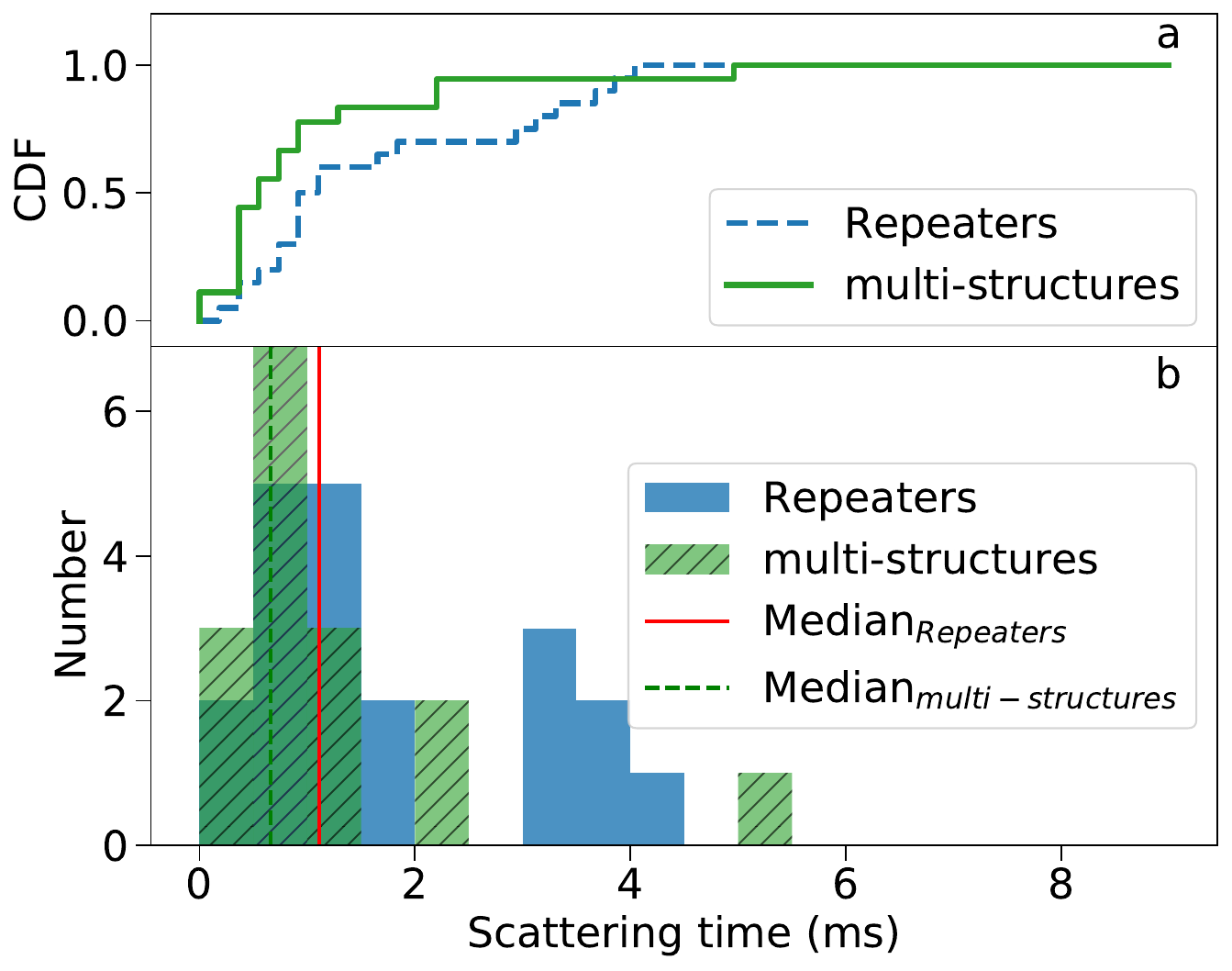}
    }
    \quad
    \subfigure[~Bandwidth distributions between repeaters and multi-structural one-offs.]{
    \label{fig:6}
    \includegraphics[width=0.45\linewidth]{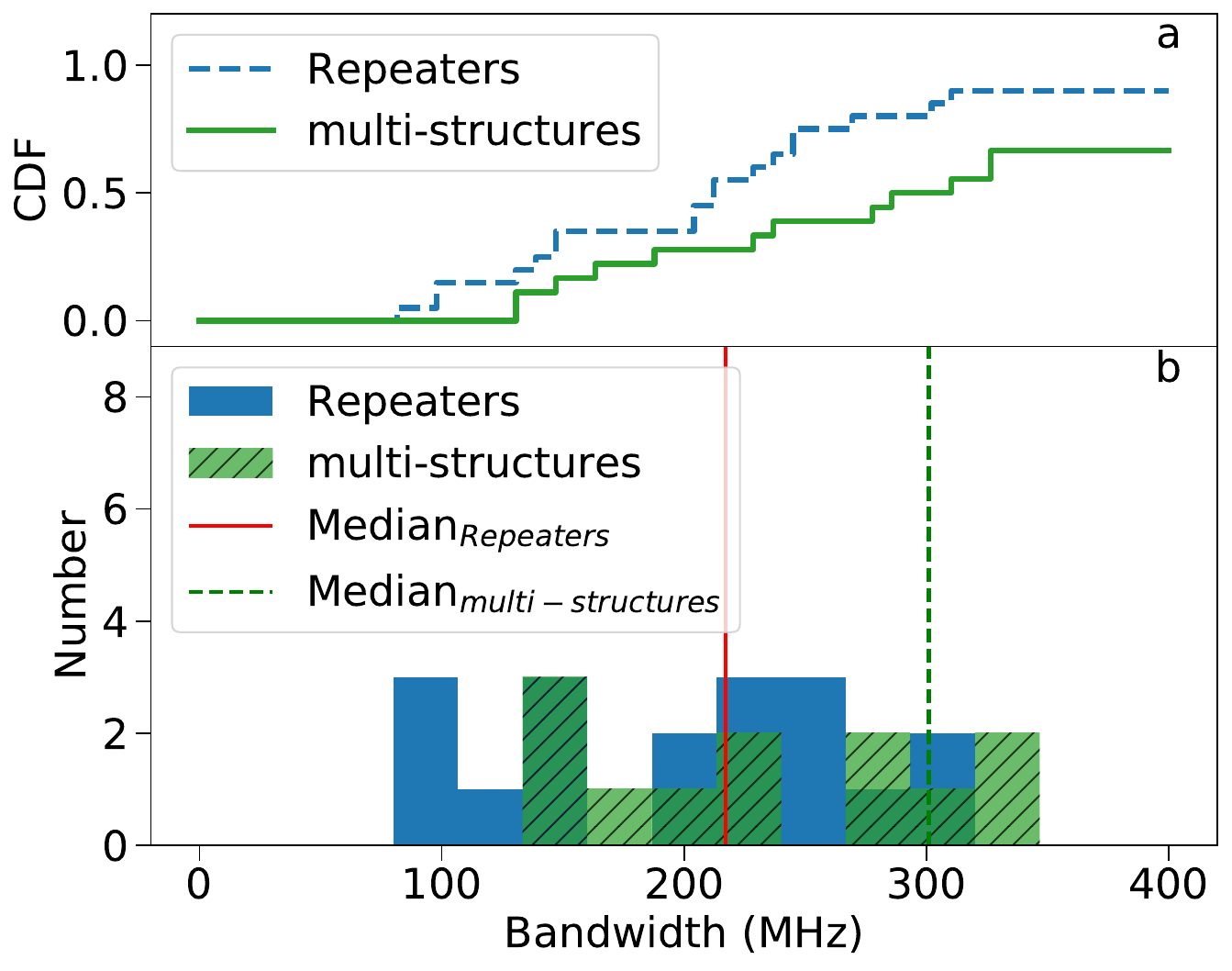}
    }
    \caption{\textls[-15]{Parameter distributions of one-offs, repeaters, and multi-structural one-offs. Here, multi-structures in (\textbf{d})--(\textbf{f}) stand for groups of multi-structural one-offs. The upper pannel, (\textbf{a}), 
gives the cumulative distributions of the observed parameters, as the lower pannel, (\textbf{b}), demonstrates the differential distributions of the observed parameters of different groups of FRBs.}}
    \label{fig:test}
\end{figure}

\newpage
\begin{adjustwidth}{-\extralength}{0cm}

\reftitle{References}

\PublishersNote{}
\end{adjustwidth}
\end{document}